\documentclass[aps,pre,twocolumn, superscriptaddress]{revtex4-1}
\usepackage{natbib}
\usepackage{amsmath}
\usepackage{graphicx}
\usepackage{nameref}

\usepackage{bm}%
\usepackage[colorlinks=true,linkcolor=blue]{hyperref}%
\usepackage{ulem}
\usepackage{changepage}
\usepackage{setspace}
\def\url#1{\textcolor{blue}{\protect\small\sf #1}}
\usepackage{color}
\newcommand {\Wm}{W m$^{- 2}~~$}
\renewcommand{\deg}{\mbox{$^\circ$}}
\newcommand {\lt}{\left(}
\newcommand {\rt}{\right)}
\newcommand{\St}{\mathcal{S}}

\begin{document}

\title{On the nature of the sea ice albedo feedback in simple models}

%
%


\author{W. Moon}
\email[]{wm275@damtp.cam.ac.uk}
\affiliation{Institute of Theoretical Geophysics, Department of Applied Mathematics  \& Theoretical Physics, University of Cambridge, Cambridge CB3 0WA, UK}
\affiliation{Yale University, New Haven, CT, 06520, USA}

\author{J. S. Wettlaufer}
\email[]{john.wettlaufer@yale.edu}
\affiliation{Yale University, New Haven, CT, 06520, USA}
\affiliation{Mathematical Institute, University of Oxford, Oxford OX2 6GG, UK}
\affiliation{Nordic Institute for Theoretical Physics (NORDITA), 10691 Stockholm, Sweden}

%
%

\begin{abstract}

We examine the nature of the ice-albedo feedback in a long standing approach used in the dynamic-thermodynamic modeling of sea ice.  The central issue examined is how the evolution of the ice area is treated when modeling a partial ice cover using a two-category-thickness scheme; thin sea ice and open water in one category and ``thick'' sea ice in the second.  The problem with the scheme is that the area-evolution is handled in a manner that violates the basic rules of calculus, which leads to a neglected area-evolution term that is equivalent to neglecting a leading-order latent heat flux.  We demonstrate the consequences by constructing energy balance models with a fractional ice cover and studying them under the influence of increased radiative forcing.  It is shown that the neglected flux is particularly important in a decaying ice cover approaching the transitions to seasonal or ice-free conditions.   Clearly, a mishandling of the evolution of the ice area has leading-order effects on the ice-albedo feedback.  Accordingly, it may be of considerable importance to re-examine the relevant climate model schemes and to begin the process of converting them to fully resolve the sea ice thickness distribution in a manner such as remapping, which does not in principle suffer from the pathology we describe.

\end{abstract}

%
%

\maketitle 

\section{Introduction\label{sec:intro}}

The original thermodynamic theory coupling sea ice and climate dealt with the system as a column of atmosphere, ice and ocean  \citep{maykut1971}.  
This approach is the cornerstone of contemporary theoretical studies \citep[e.g.,][and refs. therein]{Eisenman2012} 
and it underlies the thermodynamics of the sea ice components of all contemporary climate models \citep[e.g.,][]{AMIP}.  
We understand that the ice cover presents a distribution of ice thicknesses $g(h)$ to the atmosphere 
and ocean that force its growth, decay and deformation \citep{Thorndike1975}.  
Although the treatment of this distribution as a continuous differentiable function is based on clear reasoning, 
its practical implementation in either simple or complex models is a major challenge.  
\citet{Hibler:79} (\textit{H79} throughout this paper) developed an implementation scheme for the theory of  \citet{Thorndike1975} 
wherein both ice and open water are considered as part of a grid cell.  
In such a so-called two-category-thickness scheme, one category consists of thin sea ice and open water and the other category is ``thick'' sea ice.   The areal fraction of both categories is computed at each time step.  The scheme emerged at a time when the perennial ice state was not questioned.  
However, it does not conserve the total area of ice in a grid cell and hence, due to the nature of the ice-albedo feedback, is 
of particular importance as the state of the ice cover changes from perennial to seasonal.  
Here, we demonstrate this in a simple model.  The question of how, and how rapidly, the ice-cover may decay towards the seasonal state
is the main implication of the analysis that follows. 

It is important to note that a substantial literature on the simulation of fully resolved sea ice thickness distributions $g(h)$ began about twenty years ago \citep[e.g., ][]{Flato:1995}, an important approach being the application of the \citet{Remapping} remapping scheme to $g(h)$ by \citet{Lipscomb}.  Such approaches are not in principle influenced by the particular problem we discuss that is associated with a two-category-thickness scheme.  
However, models that continue to use a two-category-thickness scheme, or any area-thickness scheme \citep[e.g., 15 of 29 models in the Arctic Ocean Model Intercomparison Project;][]{AMIP}, could in principle be influenced by the issues we examine here.

\subsection{Multiple Sea Ice Cover States\label{sec:multiple}}

A main focus of the attempt to discern the origin of the decline of the Arctic sea ice cover is the evolution of the summer sea ice minimum \citep[e.g.,][]{Kwok:2011} and the associated question of whether future summers will be ice-free, so that there is ice only in winter.
The approaches to the problem range from theoretical treatments \citep[e.g.,][]{thorndike1992, EW09, Renate:2011, Abbot:2011, MW12, Eisenman2012, Bjork:2012, MW2013} 
and global climate model simulations \citep[e.g.,][]{holland2006, Winton:2008, Tietsche:2011}, 
to interpretation of observations \citep[e.g.,][]{Serreze:2011, Stroeve:2012, AMW:2012}.  

The rudiments of the ice-albedo feedback provide the framework for examining the nature of transitions 
from the perennial ice state to either a seasonal or ice-free sea ice state.  
In the framework of simplified versions of the column model of \citet{maykut1971} 
the ice-albedo feedback treats the sea ice albedo as a function of ice thickness $h$, 
transitioning continuously from that of sea ice to that of the ocean \citep{EW09,  Eisenman2012, MW12}.  
As the greenhouse gas forcing (modeled as an additional surface heat flux $\Delta F_0$) increases, these theoretical approaches 
capture the nature and general conditions of the transitions between, perennial, seasonal and ice-free states.  
\citet{Eisenman2012} provides a recent summary of the models and methods used to predict four general scenarios 
under which ice retreat may occur as $\Delta F_0$ increases.

\section{Partial Ice Cover \& the Ice-Albedo Feedback\label{sec:partial}}

\subsection{Column Models\label{sec:column}}

Due to the strength of the ice-albedo feedback even in the simplest of models it is important to attempt to model partial ice cover, 
which requires an ocean mixed layer that is in communication with the atmosphere unless the ocean is completely ice-covered.  
The most common two-category-thickness methodology for ice area $A$ evolution appears to have originated from \citet{Hibler:79}, discussed in more detail in \S \ref{sec:mass}.  Here, we take a minimalist approach to demonstrate the key matters at hand, which we do by implementing the \textit{H79} scheme in the simple column model of \citet{EW09}, which is derived from that of \cite{maykut1971}.  We first summarize the relevant aspects of \citet{EW09} and \citet{Ian:GFD}, and then in  \S \ref{sec:partial1} we describe the implementation of \textit{H79}.  

When the temperature of the ice $T_i <  0$ \deg C, it evolves in time $t$ along with the ice thickness $h$ according to 
\begin{align}
    \frac{c_{pi} h}{2} \frac{d T_i}{d t}&= F_{top} - k T_i/h  \qquad \text{and}
  \label{eq:T92-freeze1} \\
    L\frac{d h}{dt}&=-k\frac{T_i}{h}-F_{B} , 
  \label{eq:T92-freeze2}
\end{align}
where $L$ is the latent heat of fusion per unit volume, $c_{pi}$ is the specific heat capacity of ice at constant pressure, and $k$ is the thermal conductivity.   
The sum of sensible, latent, downward and upward longwave, and shortwave heat fluxes at the surface is given as $F_{top}$, 
and all but the upward longwave flux are specified from observed radiation climatology as in \citet[][]{EW09}. 
The flux from the ocean mixed layer into the base of  the ice is $F_{B}$.

When $T_i$ = 0 \deg C the temperature evolves along with the ice thickness according to 
\begin{align}
  \frac{d T_i}{d t}&= 0 \label{eq:T92-melt1} ,\\
  L\frac{d h}{dt}&=-F_{top}-F_{B} .
  \label{eq:T92-melt2}
\end{align}
The ocean mixed layer is treated as a thermodynamic reservoir with a typical observationally-based characteristic depth of $H_{ml}=50$m and heat flux entrained through the bottom of the mixed layer of $F_{ent}=0.5$ \Wm. 
The turbulent heat flux between the ocean and the underside of the ice is a complex quantity modeled crudely here as being 
proportional to the elevation of the mixed layer temperature $T_{ml}$ above freezing by $F_B=\rho_w c_{pw} c_h u_{\star \text{o}}T_{ml}$, in which $\rho_w$ and $c_{pw}$ are the density and specific heat capacity at constant pressure of seawater, $c_h=0.006$ is the heat transfer coefficient, and $u_{\star \text{o}}$ is the square root of
kinematic stress at ice-ocean interface, also known as the friction velocity \cite[e.g.,][]{MaykutMcPhee}.  A typical observational value of
$u_{\star \text{o}}=0.5$~cm s$^{-1}$ leads to $F_B=\gamma T_{ml}$ with $\gamma\equiv\rho_w c_{pw} c_h u_{*0} =120 \text{Wm}^{-2}/\text{K}$.
Measurements show that while in the upper summer mixed layer $T_{ml}$ can be as much as 0.4 \deg C above freezing, it is an order of magnitude smaller in winter, giving a seasonally averaged $F_B$ of about 5 \Wm~\cite[][]{MaykutMcPhee}.  Therefore, according to whether $T_i<0$\deg C, a continuously evolving seasonal cycle is captured, 
using \eqref{eq:T92-freeze1}-\eqref{eq:T92-freeze2} or \eqref{eq:T92-melt1}-\eqref{eq:T92-melt2}. 

\subsection{Modeling Partial Sea Ice Cover} \label{sec:partial1}

Now we proceed to the issue of modeling partial ice cover using column models.   
We begin by summarizing the commonly used approach to this problem developed by \citet{Hibler:79}.  
Such a method can be rationalized physically for the perennial ice cover for which it was developed,  but we discuss its behavior when the ice fraction decreases, 
such as is relevant during the transition to seasonal ice as the observed state of the ice cover changes \citep{Perovich:2012}. 

\begin{figure}[h!]
\noindent
\includegraphics[width=18pc, height=13pc]{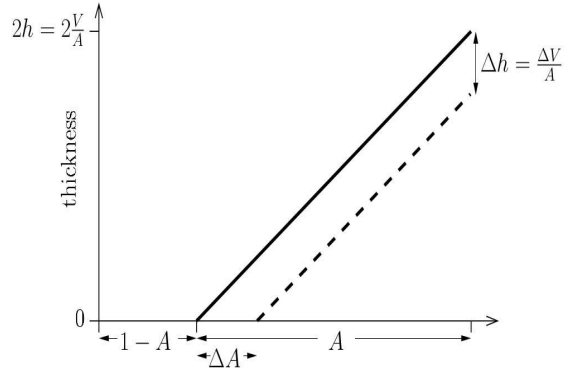}
  \caption{Schematic illustrating the proportionality between the rate
    of change of ice area $A$ and the thermodynamic decrease of volume following \cite{Hibler:79}.}
\label{fig:AreaSchem}
\end{figure}

The \textit{H79} methodology to determine ice concentration $A$, or the fraction of a grid cell covered by ice, is the core focus here.
This requires a form of homogenization over the subgrid-scale to account for open water.  As noted in the introduction, 
this method provides the framework for parameterizations in two-category-thickness distributions for area-thickness modeling schemes.  Of the 29 models participating in the Arctic Ocean Model Intercomparison Project \citep[][]{AMIP} 15 use area-thickness schemes. For clarity, but without loss of generality, we discuss the \textit{H79} approach in terms of a model that includes a single grid cell.  The approach applies to either the area of ice in the grid cell or, as is done in \textit{H79}, the fraction of the grid cell covered by ice $A$.  Although in many of the equations that follow these can be used interchangeably, we use areal fraction for consistency with \textit{H79}.  This means that in the ice-covered fraction of the grid cell ice thickness $h$ becomes the volume per area $V$, which has units of length.  Variables and constants are defined in Table~\ref{table:parameters}.

The ice concentration increases when $T_{ml}$ reaches zero and continues to cool so that the mixed layer flux imbalance $F_{ni}$ 
drives the creation of new ice as 
\begin{equation}
\frac{d A}{d t}=\frac{F_{ni}}{L h_0} .
\end{equation}
An ``equivalent thickness'' $h_0$ is assigned to the new area ascribing volume to it. 
Thus, area increases only when the mixed layer freezes, but once it does so, the new volume of that ice increases only
by increasing the ice thickness at fixed area. Because, within the framework of column models, 
sea ice growth rate is calculated (or specified) as a function of ice thickness and season, 
the value of $h_0$ controls the rate at which the ice cover grows.  The  value of $h_0$ used in \textit{H79} is 50 cm. 
Although the growth rate in winter decreases by a factor of four as open water solidifies to a thickness of 50 cm, 
the ice concentration in \textit{H79} increases based on the growth rate for open water $\sim$ 12 cm day$^{- 1}$ 
(see figure 3 in \textit{H79}). 
Importantly, in this and similar two-category-models, the \textit{open water fraction} is not meant to represent an entirely ice-free region. 
Rather, the model domain is split into a fraction containing thick ice, with the rest covered by a mixture of open water and thin ice, 
such as in leads. The volume of this thin ice is assumed to be negligible compared to the thick ice volume, 
which as we shall see in \S \ref{sec:mass} is one of the problems in dealing quantitatively with processes such as ice-albedo feedback. 

Energy balance dictates that area decays in this model when volume ablates ($\frac{d V}{d t}<0$) and hence 
\begin{equation}
\frac{d A}{d t}=\frac{A}{2 V}\frac{d V}{d t}.
\end{equation}
The proportionality between volume and area rates of change is based
on an argument about the ice thickness distribution in the
model domain under the following assumptions \citep{Hibler:79}: (a) the ice is linearly distributed in thickness
between 0 and $2 V/A$, thereby giving a mean thickness of $V/A$, and (b) all of this ice melts at
the same rate. As illustrated schematically in Fig.~\ref{fig:AreaSchem}, this gives a rate of area decay as the rate of
thickness decay times the inverse slope of the thickness distribution; 
\begin{equation}
  \Delta A=\Delta h \frac{dA}{dh}=\frac{\Delta V}{A}\frac{A}{2V/A}=\frac{A}{2V}\Delta V.
  \end{equation}
We note here that ice growth is a nonlinear function of thickness and
here it is computed under the assumption that all ice within $A$ is of the mean thickness 
$V/A$ as opposed to the linear distribution between 0 and $2 V/A$ used for ablation.

\begin{figure}[h!]
\noindent
\includegraphics[width=14pc, height=14pc]{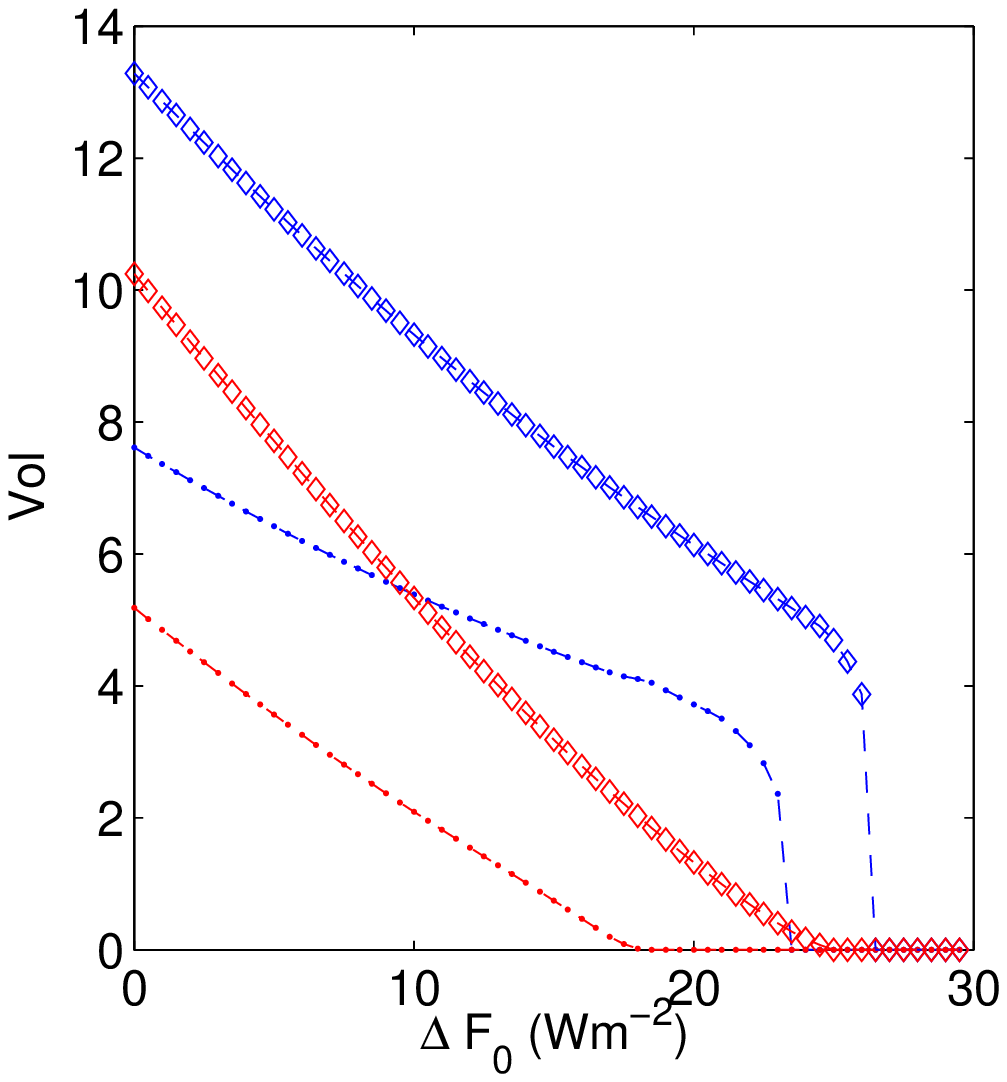}\\
\includegraphics[width=14pc, height=14pc]{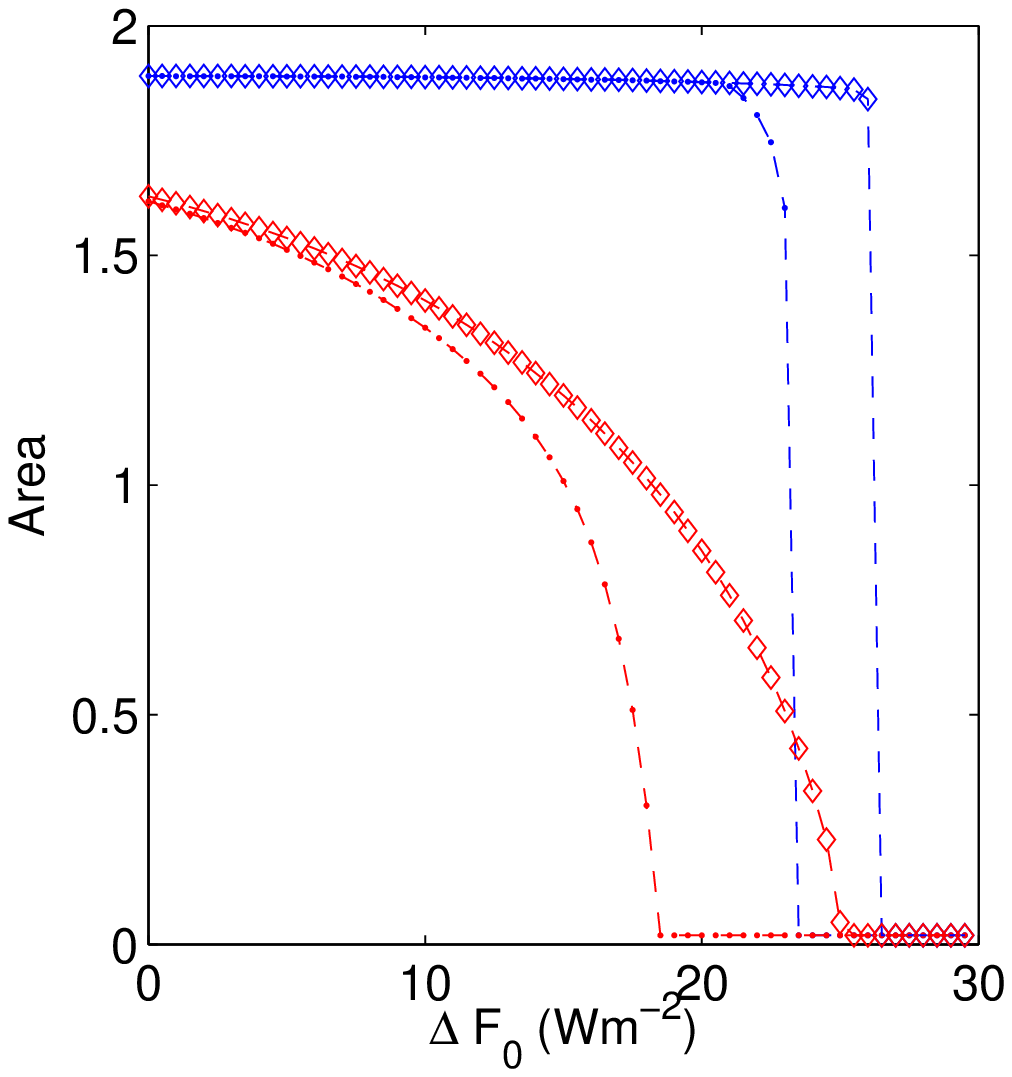}\\
\includegraphics[width=14pc, height=14pc]{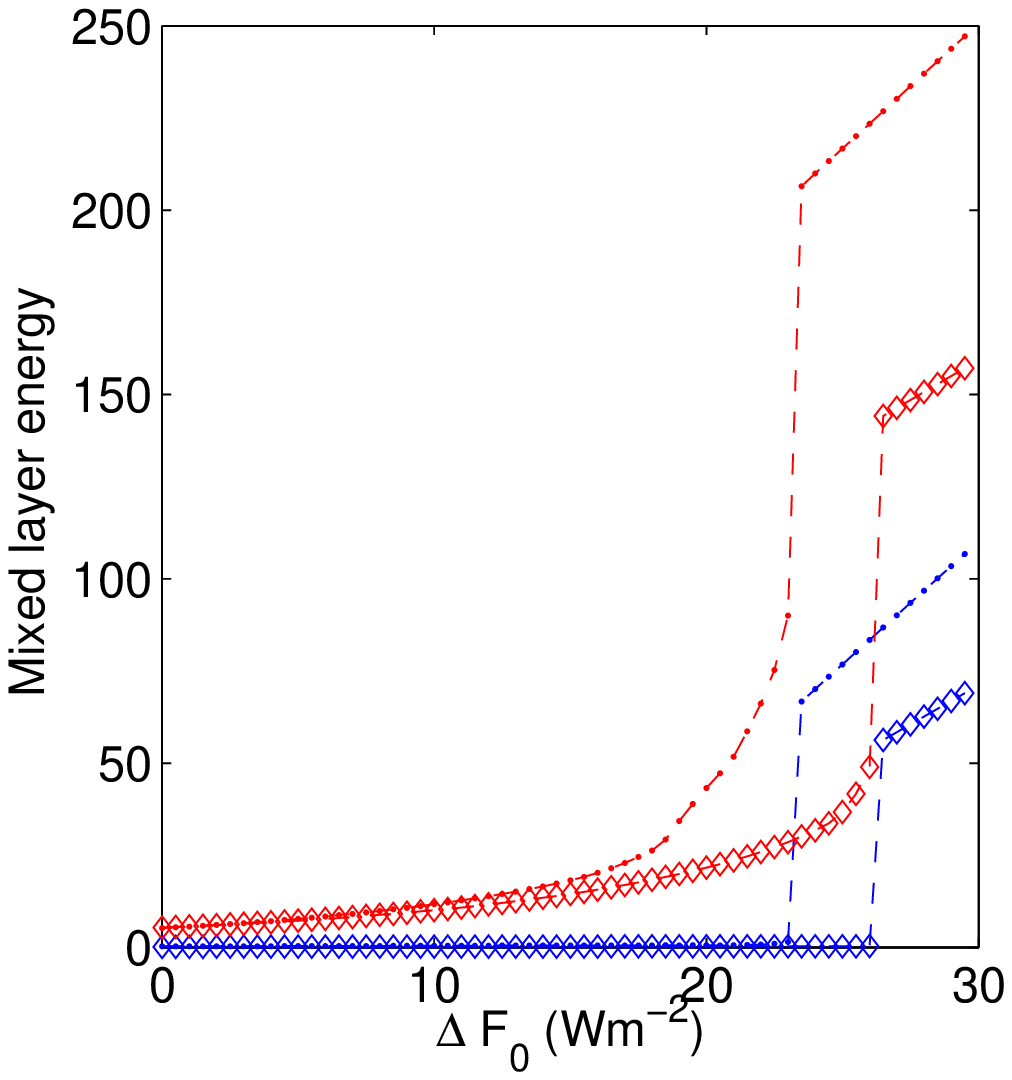}
\caption{The three panels show two sets of solutions for the evolution of the dimensionless ice volume $\cal{V}$, 
         ice area $\cal{A}$, and the mixed layer energy as a function of greenhouse gas forcing $\Delta F_0$ (W m$^{- 2}$).  
         The {\it dash-dot lines} ({\it open diamonds}) show the solutions of the equations above 
         in which {\it area evolution is complete} ({\it area evolution is incomplete}).  
         In both cases red shows the end of the summer (late August) and blue the end of winter (late March).  
         The differences are discussed in the text in more detail, but we note here that a principal feature is 
         that when area evolution is incomplete, substantially higher values of $\Delta F_0$ are required before seasonal or perennial ice is lost.}
      \label{fig:Results}
\end{figure}

Finally, the persistent convergence and divergence of the wind field results in an observed net average annual export of $v_0$ = 10\% of the ice area.  
Thus, the ice dynamics are represented in such a model by requiring that $A\leq0.95$, and a term $-v_0 A$ is added to the area-evolution equation 
(which accounts for volume export).  Export is included in the results shown figure \ref{fig:Results}, but to avoid the clutter in the theoretical development  
we omit the term in the equations that follow because it has no effect on the main points. 

Using such a scheme one can derive a partial ice cover model from the column treatment of \S \ref{sec:column} as follows. 
We determine ice volume rather than ice thickness. In the ice-covered fraction of the model domain
$A$, the vertical thermodynamic growth of the ice is represented by
re-writing \eqref{eq:T92-freeze2} and \eqref{eq:T92-melt2} as 
\begin{align}
  L\frac{d V}{dt}=A \lt -k\frac{T_i}{h}-F_{B} \rt,
  \label{eq:T92-freeze3}
\end{align}
and
\begin{align}
  L\frac{d V}{dt}=A \lt -F_{top}-F_{B} \rt .
  \label{eq:T92-melt3}
\end{align}
The total heat flux into the mixed layer is written as
\begin{align}
  F_{ml}=&- A
  \gamma T_{ml}+F_{ent} \nonumber \\
  + & (1-A)\left[-F_{lw}(T_{ml})+(1-\alpha_{ml})F_{sw}\right] ,
    \label{eq:Fml}
\end{align}
where $F_{lw}(T_{ml})$ is the net surface longwave radiation balance which depends on the mixed layer temperature $T_{ml}$, 
the shortwave radiative balance is $F_{sw}$, and the albedo of the mixed layer is $\alpha_{ml}$ \citep{EW09, Ian:GFD}.  
Therefore, if $T_{ml}>0$, this leads to heating or cooling according to
\begin{equation}
c_{pw}H_{ml}\frac{d T_{ml}}{d t}=F_{ml} ,
\end{equation}
and no new ice area is formed, $F_{ni}=0$. However, when the mixed layer
reaches the freezing temperature ($T_{ml}=0$), supercooling is prohibited such that $d T_{ml}/d
t=0$, and any additional heat loss is available to form new ice ($F_{ni}=-F_{ml}$).

\subsection{Area Evolution and Ice-Albedo Feedback\label{sec:mass}}

The principal issue here is that Eqs. \ref{eq:T92-freeze3} and \ref{eq:T92-melt3} do not correctly capture 
the evolution of the areal fraction of ice.  The nub of the matter is the appropriate grid homogenization of the mixture theory.  
We use the logic of \textit{H79}, 
that there are two ice categories, with ``thick ice'' (that we call $h_i$) covering an area fraction $A$ of a grid cell and ``thin ice'', up to a thickness $h_0$, covering the remaining fraction of the grid cell $1-A$.
\textit{H79} states that ``since the thin ice mass is normally small, the mean thickness of the remaining `thick' ice is approximately equal to $h/A$.''

The core of the thermodynamic component of Hibler's model is described by his equations (13) and (14), which take the forms
\begin{equation}
  {\partial h\over\partial t} = A f(h/A) + (1-A) f(0) {\text{~~~and}}
  \label{H79hevolution}
  \end{equation}
  \begin{equation}
  h_0{\partial A\over\partial t} = (1-A)f(0)  
   \label{H79Aevolution}
\end{equation}
during growth, for example.

Hibler does not close his mathematical description with an expression for the functions $f(h/A)$ and $f(0)$, rather defining in words that $f(h)$ is ``the growth rate of ice of thickness $h$'', which is ``based on the heat budget calculations by \citet{maykut1971}''.  A natural interpretation of this language is that
\begin{equation}
   {\partial h_i\over\partial t} = f(h_i)=f(h/A).  
 \label{hievolution}
\end{equation}
Because $h=h_iA$ holds for all time, by application of the product rule to $\partial(Ah_i) / \partial t$, equations (\ref{H79hevolution})--(\ref{hievolution}) are inconsistent unless $h_0=h_i$, which is generally not taken to be the case in \textit{H79}.  A more troublesome set of inconsistencies occurs during melting, which, although one has been noted previously \citep[e.g.,][]{Ritz:2011}, are too involved to describe here.  Hence, we simply adopt the \textit{H79} melting scheme without question in \S \ref{sec:partial1} for demonstration purposes. 

Although the \textit{H79} growth scheme is often cited as the two-category-thickness approach that is used, strictly speaking, what is implemented may only be heuristically related to the original scheme.  For example, in \textit{H79} the sea ice sits upon a reservoir--the ocean--fixed at the freezing point.  In contrast, in one model that cites implementation of the \textit{H79} growth scheme an energy balance is assessed in an upper ocean layer, and there is no explicit equation for the mean thickness of \textit{H79} as in equation (\ref{H79hevolution}) \citep[see equation (30) of][]{Marsland:2003}.  It is impractical to attempt to detail all such examples. It is sufficient to note that if an interpretation of Hibler's scheme is to avoid the inconsistency of equations (\ref{H79hevolution})--(\ref{hievolution}) described above, then it must not mathematically equate $f(h)$ with the growth rate of ice of thickness $h$ as described by equation (\ref{hievolution}).

The appropriate conservation law requires the addition of the term $ - L \frac{V}{A} \frac{d A}{dt}$ to the left hand side of Eqs. 
\ref{eq:T92-freeze3} and \ref{eq:T92-melt3}.  
To assess the importance of such a term and facilitate simple analysis and interpretation of general features of the freezing and melting process,  it is prudent to render these equations dimensionless through the introduction of the following scalings; 
\begin{align} 
V&={\cal V} V_0,  &  A&={\cal A} A_0,& T_i &= {\cal T}_i \Delta T,  \nonumber \\
t &={\cal T} \tau,& \St&=\frac{L}{c_{pi} \Delta T}, &F_c&=k \frac{\Delta T}{h_0}, 
\label{eq:params} 
\end{align}
where $h_0=\frac{V_0}{A_0}$ is the threshold thickness mentioned above, $c_{pi}$ is again the specific heat capacity of the ice, 
$\Delta T$ is the temperature difference over ice of thickness $h_0$ and hence $F_c$ is the associated conductive heat flux.  
The dimensionless ratio $\St$ is a Stefan number, which represents the relative importance of latent heat to the specific heat in the ice, 
and is large ($ > 10$) here.  
These scalings lead to dimensionless versions of Eqs. \ref{eq:T92-freeze3} and \ref{eq:T92-melt3} appropriately modified to include the area evolution as
\begin{align}
  \St \left[\frac{d {\cal V}}{d \cal T} - \frac{\cal V}{\cal A} \frac{d {\cal A}}{d{\cal T}}\right]={\cal A} \lt -{{\cal T}_i}\frac{\cal A}{\cal V}-{\cal F}_{B} \rt,
  \label{eq:newfreeze}
\end{align}
and
\begin{align}
  \St \left[\frac{d {\cal V}}{d \cal T} - \frac{\cal V}{\cal A} \frac{d {\cal A}}{d{\cal T}}\right]={\cal A} \lt -{\cal F}_{top}-{\cal F}_{B} \rt,
  \label{eq:newmelt}
\end{align}
where the fluxes $\cal F$ are just the dimensional fluxes scaled by $F_c$.  
Because $\St \gg 1$, the balance in both Eqs. \ref{eq:newfreeze} and \ref{eq:newmelt} requires that 
\begin{align}
\left[\frac{d {\cal V}}{d \cal T} - \frac{\cal V}{\cal A} \frac{d {\cal A}}{d{\cal T}}\right] \ll 1,   \nonumber \\
\Longleftrightarrow \frac{d {\cal V}}{\cal V}  \sim \frac{d {\cal A}}{{\cal A}}, 
 \label{eq:neglect}
\end{align}
showing that the neglected term is of the same order of magnitude as that kept in the scheme.  
Importantly, in transitioning to a seasonal ice state driven by the ice-albedo feedback, 
correctly capturing the rapidly changing evolution of ice area $\frac{d {\cal A}}{d{\cal T}}$ is crucial, otherwise it is found that 
ice loss is only expected in untoward parameter regimes.
Indeed, when \citet{Ian:GFD} neglected this term and used the correct value of the latent heat of fusion of ice he could not simulate a realistic ice cover.  
For example, he found that in order to obtain multiple sea ice states under greenhouse gas forcing of 1.5 times the present value, he had to artificially decrease $L$ by factors ranging from four to ten.  This reflects the fact that, for a given radiation balance, the artificially large latent heat flux associated 
with the neglect of the extra term in Eqs. \ref{eq:newfreeze} and \ref{eq:newmelt} can only be balanced by positing an artificially low value of $L$.  
Moreover, when using the correct thermophysical constants, he had to use 5 times (16 times) 
the present greenhouse gas concentration to transition from perennial to seasonal ice (multiple sea ice states). 

\subsection{Energy Flux Conservation}

Having demonstrated the size of the missing term, we return to dimensional variables in this section.  During the melt season the contribution to the volume evolution of the lateral melting term $h\frac{dA}{dt}$ 
 is insufficient to conserve energy flux.  In the \textit{H79}
 scheme, this lateral melting is calculated indirectly in order to maintain the functional form of the model sea ice thickness
 distribution, save for the constant difference owing to the change of the mean.  
 Because greater than 90\% of the incident sunlight is absorbed 
 by open water, we understand that the partitioning of the ablation of the ice 
 cover between top, bottom and lateral boundaries is, among other factors, a complex function of the open water fraction 
 and ice floe perimeter \cite[e.g.,][]{Maykut:1987, Steele:1992}. The impasse faced by the \textit{H79} approach 
is discussed by \citet{Steele:1992} in terms of the lack of an explicit equation for ice floe perimeter.  
It is thus natural to ask for the origin of the heat source for lateral melting within this model framework.  

We examine the partitioning of vertical and lateral oceanic-heat fluxes to account for the contribution $h\frac{dA}{dt}$.  
When the average thickness $\overline{h}$ is used to determine the heat flux required to balance the volume
change originating in vertical ablation $A\frac{dh}{dt}$, the result differs from the analogous procedure 
in which the ice thickness is distributed evenly from $0$ to $2\overline{h}$.  Thus, we
conclude that part of this heat flux difference $\Delta F_i$ over an ice area $A$ is used for lateral melting.  Hence, if $F_i(h)$ is the net heat flux over sea ice of thickness $h$, we write $\Delta F_i$ as 
 \begin{align}
   \Delta F_i &= \int_{0}^{2\overline{h}}F_i(h)g(h)dh - AF_i(\overline{h}),   \text{~~where~~}\\
  g(h)&=\frac{A}{2\overline{h}}, \text{~~and thus~~}\\
   A&=\int_{0}^{2\overline{h}}g(h)dh \text{~~and~~} g(0)=1-A.
 \end{align}
For simplicity of exposition, we view the principal contribution to $\Delta F_i$ as arising from shortwave radiative fluxes as
  \begin{align}
  \Delta F_i &= \int_{0}^{2\overline{h}}[1-\alpha(h)]F_{sw}(t)g(h)dh - A[1-\alpha(\overline{h})]F_{sw}(t) \nonumber \\
             &=\frac{\alpha_i-\alpha_{ml}}{2}F_{sw}(t) 
             A\left\{-\frac{h_0}{2\overline{h}}\text{ln}
             \left[\text{cosh}\left(\frac{2\overline{h}}{h_0}\right)\right]+\text{tanh}\left(\frac{\overline{h}}
              {h_0}\right)\right\}.
 \end{align}
Therefore, the lateral oceanic-heat flux ablating the ice is $L h\frac{dA}{dt}-\Delta F_i$. Finally, in order to conserve  
 energy flux balance at each time step during the melt season, we subtract the lateral oceanic-heat flux from the evolution of
 ocean sensible heat, thereby avoiding an anomalous increase in ocean heat content. 
 
\section{Discussion}

Now that the essential point has been made using this simple analysis, 
in Figure \ref{fig:Results} we show the dramatic effect of employing a scheme that deals with the area evolution as discussed above.  
The missing-area (and hence ice-mass) term, as described in the argument leading to equation \ref{eq:neglect}, 
has the basic effect of neglecting a leading-order latent heat flux in the energy balance; because $\frac{d {\cal A}}{d{\cal T}}$ and $\frac{d {\cal V}}{d{\cal T}}$ have 
the same sign, not including the missing-area leads to a larger latent heat flux.  
Under the same radiative forcing the consequences of this missing latent heat flux can be clearly laid bare.   
Firstly, because the effective latent heat flux is larger than it should be, the volume of ice in steady state 
with the same radiative balance can be larger as seen in the top panel.  
The most distinct case is for $\Delta F_0$ = 0, where the maximum sea ice thickness is 3.5 m (2 m) 
when this latent heat flux is ignored (included).  
Secondly, when this latent heat flux (and hence the associated ice areal change) is ignored, a larger value of $\Delta F_0$ ($\sim$ 5 W m$^{- 2}$) 
is required before both the perennial and seasonal ice states vanish.   
Thirdly, the range of $\Delta F_0$ over which seasonal sea ice exists in a stable state is infinitesimal (practically non-existent) when the
area evolution is incomplete whereas it is $\sim$ 5 W m$^{- 2}$ when the area evolution and hence latent heat flux is complete. 
Finally, since the ice cover vanishes at smaller values of greenhouse forcing when area evolution is treated completely, and hence the ice-albedo feedback is appropriately captured, the heat content of the exposed mixed layer is larger.  
We note that in the original \textit{H79} treatment there was no mixed layer; 
the ocean temperature was constrained to lie on the freezing point and any excess heat absorbed was immediately 
added to a basal heat flux and applied to the underside of the thick ice.  
This treatment is clearly unrealistic in the limit of a vanishing ice cover when
the missing term in such a scenario becomes all the more important, 
and a direct application of this scheme simply amplifies the differences shown in figure \ref{fig:Results},
so we do not include these figures here. 

It is important to note that although we have focused on the simplest (two-category-thickness) schemes, our arguments can be generalized. In a multi-thickness model, $A=\int_{0}^{h_{max}}g(h)dh$ and $1-A = g(0)$. Hence, regardless of the scheme, one must provide some form of rule for $1-A$,  which may for example include the 
lateral heat flux to sea ice. Because one must construct some rule to evolve the open water or thin ice fraction, our main point is general, whereas using a remapping scheme \cite[][]{Remapping} to solve a thickness distribution equation \citep[][]{Lipscomb} is in principle free from this problem.

\section{Conclusion}

We describe how a long standing approach used in the thermodynamic modeling of sea ice does not treat the complete evolution of the ice area and thus cannot capture the influence of the ice-albedo feedback.  The missing-area term, as described in the argument leading to equation \ref{eq:neglect}, 
has the effect of neglecting a leading-order latent heat flux in the energy balance.  By deriving energy balance models for partial ice cover with and without the appropriate area evolution we have demonstrated the sensitivity of the results 
to the missing area. It is found to be particularly important in a decaying ice cover approaching seasonally ice-free conditions.  
Although we have not independently analyzed how this erroneous treatment of area evolution has propagated through the range of GCMs used, 
our analysis indicates the possibility that it could in fact be one of the underlying features responsible 
for the observed recent Arctic sea ice decline being more rapid than is forecast using a multi-model ensemble mean \citep[e.g.,][]{Stroeve:2012}.   Thus, it is suggested that it may be of considerable importance
to re-examine the relevant climate model schemes and to begin the process of converting them to fully resolve the sea ice thickness distribution in a manner such as remapping \cite[][]{Remapping, Lipscomb}, which does not in principle suffer from the pathology we have described here.


%
%
%
%
%
%

%
%
%
%
\vspace{-0.75cm}
\begin{acknowledgments}
W.M. thanks NASA for a graduate fellowship and J.S.W. thanks the John Simon Guggenheim
Foundation, the Swedish Research Council, and a Royal Society Wolfson Research Merit Award for
support.   The authors thank Ian Eisenman, Georgy Manucharyan, George Veronis  and Grae Worster for a series of educational exchanges during the evolution of this project.    We also thank D. Notz who stimulated their re-assessment of the potential range of interpretations of the \citet{Hibler:79} scheme. We note that there are no data sharing issues since all of the numerical information is provided in the figures produced by solving the equations in the paper. 
\end{acknowledgments}

%

\eject
{\setstretch{0.75}
\begin{table}[ht]
\begin{adjustwidth}{-2cm}{}
 \caption{Descriptions and values of state variables and model parameters}
\begin{center}
 \begin{tabular}{l l l}
  \hline
 Symbol & Description & Units/Value \\
  \hline
\label{table:parameters}
   $h$ & Ice thickness & m \\
    $V$ & Ice volume per unit grid cell area & m \\
    $A$ & Ice areal fraction in grid cell & $0 \le A \le 1$ \\
    $T_i$ & Ice surface temperature & \deg C \\
    $T_{ml}$ & Ocean mixed layer temperature & \deg C \\
   $h_0$ & equivalent thickness for newly formed ice & 0.5 m \\
    $L$ & latent heat of fusion per unit volume & $3 \times 10^8$ J/m$^{3}$ \\
     $k$ & thermal conductivity of ice  & 2 W/m/K \\
   $\rho_{w}$ & density of water at constant pressure & $1 \times 10^3$ kg/m$^{3}$ \\
  $c_{pi}$ &  specific heat capacity of ice at constant pressure & $2 \times 10^6$ J/m$^3$/K \\
   $c_{pw}$ & specific heat capacity of water at constant pressure & $4 \times 10^6$ J/m$^{3}$/K \\
  $\alpha_i$ & albedo of ice & 0.65 \\
  $\alpha_{ml}$ & albedo of ocean mixed layer & 0.20 \\
 $F_{sw}$ & shortwave radiation at ice or ocean surface & seasonal;  \Wm \\
$F_{lw}$ & longwave radiation at ice or ocean surface & seasonal;  \Wm \\
  $F_{top}$ & net surface sensible, latent and radiative heat flux & \Wm \\
  $\Delta F_0$ & greenhouse gas forcing & 0-30 \Wm\\
$c_h$ & ocean-ice heat transfer coefficient & 0.006 \\
$u_{\star \text{o}}$ & ocean-ice friction velocity & 0.5 cm s$^{-1}$ \\
  $\gamma$  & ocean-ice heat exchange coefficient = $\rho_w c_{pw} c_h u_{*0}$ & 120 \Wm/K\\
  $H_{ml}$ & mixed layer depth & 50 m \\
   $F_{entr}$ & heat flux entrained into mixed layer base & 0.5 \Wm\\
  $F_{B}$ & seasonal average ice-ocean heat flux & 5 \Wm\\
  $F_{ml}$ & total heat flux into the mixed layer (Eq. \ref{eq:Fml}) & seasonal;  \Wm \\
    \hline
 \end{tabular}
\end{center}
\end{adjustwidth}
\end{table}
}


\end{document}